# The ringdown-Hawking radiation connection in real and analogue black holes

Eyal Keshet, Inbar Shemesh and Jeff Steinhauer

*Department of Physics, Technion – Israel Institute of Technology, Technion City, Haifa 3200003, Israel*

In the usual picture of Hawking radiation, the emission is spontaneous; it is caused by nothing. In contrast, the radiation from the ringdown after a black-hole merger is caused dynamically by the fluctuations of the event horizon. We explore the possibility that Hawking radiation is also emitted dynamically by horizon fluctuations, in the form of quasinormal modes. In fact, we find that the fundamental, least-damped quasinormal mode is sufficient to radiate the entire Hawking spectrum of photons and gravitons, since the quasinormal mode spectrum is broadened by damping. The resulting Hawking spectra are accurate with no graybody factors. By comparing Hawking radiation to ringdowns, we find that the quantum fluctuations of the horizon should be on the order of 0.1 Planck lengths. We compare this result with predictions ranging over 60 orders of magnitude. Further support for the model is provided by a sonic black hole experiment, in which additional horizon fluctuations are seen to produce Hawking-like radiation.

During the ringdown after a black hole merger [1-6], the horizon of the black hole is in a perturbed state, which results in the dynamical emission of gravitational waves via quasinormal modes [7, 8]. In contrast, Hawking emission of blackbody radiation requires no perturbation of the horizon [9]. One can compute the resulting perturbation, but it is not required *a priori*. On the other hand, a black hole is a rather unusual blackbody, since its temperature is lower than the lowest resonant mode. In such a system, one expects the usual continuous blackbody spectrum to be replaced by a series of discrete spectral lines, corresponding to the few lowest modes. We will see that this is exactly the case for Hawking radiation, where the blackbody spectrum is replaced by the fundamental quasinormal modes for photons and gravitons. This immediately leads to a dynamical description of Hawking emission, since the radiation is directly emitted by fluctuations of the horizon, in the form of quasinormal modes. Moreover, we find that only one quasinormal mode is required – the least damped, fundamental quasinormal mode has a sufficiently broad spectrum to radiate the entire Hawking spectrum computed in Ref. 10.

Stephen Hawking outlined a quantum gravitational description of Hawking radiation, by asserting that metric fluctuations should allow particles to tunnel out of a black hole [9]. James Bardeen predicted that the fluctuations in the radius of the horizon should be $\ell_\text{P}^2/R$, where $R$ is the radius of the horizon and $\ell_\text{P}$ is the Planck length [11]. The corresponding fluctuations in the area of the horizon would be $\ell_\text{P}^2$, which is the area quantum [12-15]. By invoking non-radiative quasinormal modes [16], James York found a vastly larger value of 0.02 $\ell_\text{P}$. Newtonian gravity gave an even larger prediction of $\ell_\text{P}^{2/3} R^{1/3}$ [17], and including relativistic effects further increased the estimate to $\sqrt{\ell_\text{P} R}$ [18]. These predictions vary over 60 orders of magnitude. In this work, we consider the possibility that the metric fluctuations take the form of quasinormal modes which emit Hawking radiation. Via the quasinormal modes, we can determine the amplitude of the quantum fluctuations of the horizon, by studying the classical oscillations of the ringdown. This is the opposite of works in which quantum effects modify the classical observations [19-22].

**Sonic black hole experiment**

Before studying real black holes, we would like to verify the dynamical Hawking radiation picture with a sonic black hole. Can microscopic fluctuations of the horizon generate Hawking radiation? In order to simulate such fluctuations, we will induce extra fluctuations beyond the zero-point level, and see whether the resulting radiation is similar to spontaneous Hawking radiation, but with larger amplitude.

The experimental sequence is shown in Fig. 1(a). An atomic cloud near condensation is created in a magnetic trap (not shown), which begins to ramp down at 31 s. An optical dipole trap (a red-detuned focused laser beam) is turned on at 33 s, as indicated by the magenta curve. The magnetic trap is completely off at 35.5 s, at which time the power in the optical dipole trap is ramped down, resulting in decompression and evaporative cooling to sub-nanokelvin temperatures. A blue-detuned step potential beam (black curve and Fig. 1(b)) is turned on at 35.5 s. The position of the edge of the step potential is controlled by the angle of a rotating mirror (red curve). At first, the edge of the step is well outside the atomic cloud, but at 37 s, the edge starts to move with constant subsonic speed. In the reference frame of the edge, this corresponds to motion of the atomic cloud (Fig. 1(b)) to the right. The cloud flows over the edge, forming the supersonic ($v > c_s$) and subsonic ($v < c_s$) regions of Fig. 1(b), where $v$ and $c_s$ are the flow velocity and speed of sound, respectively. For the observation of spontaneous Hawking radiation, the power of the step potential is a constant (unlike the black curve of Fig. 1(a)), and the edge moves with a constant speed (the ramp of the red curve). The correlation plot of Fig. 2(a) shows the spontaneous Hawking radiation, which is similar to the previous observations [23-25], and to the predictions for real black holes [26-28]. Specifically, the gray band parallel to the green line is the correlations between the Hawking and partner particles. The profile of this Hawking/partner band is indicated by solid curves in Figs. 2(i) and 2(j). Distances are expressed in units of the healing length $\xi$.

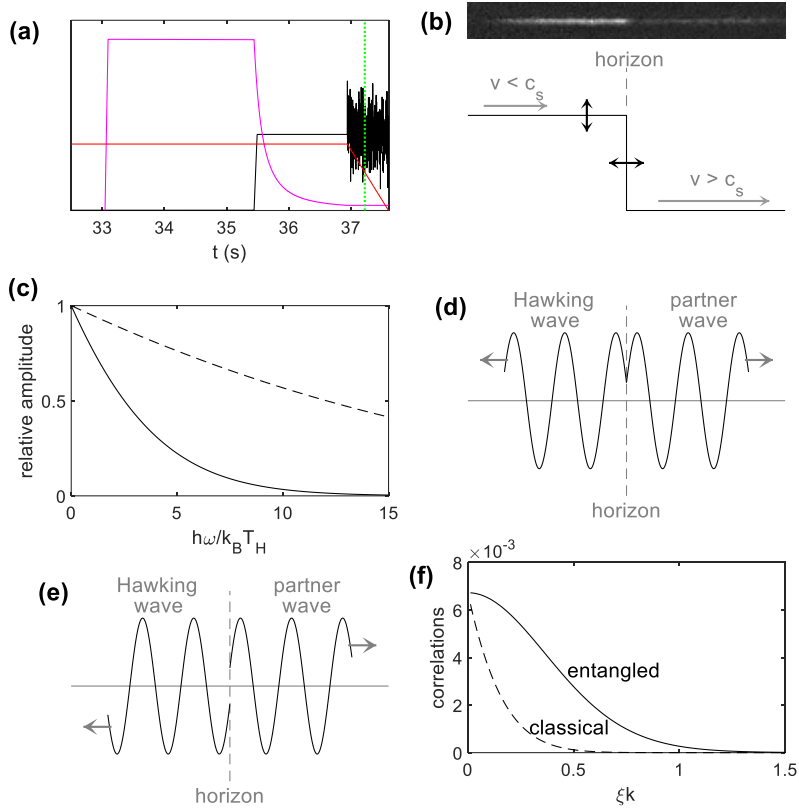

**Figure 1. Experimental technique.** (a) Timing diagram. The magenta curve is the power of the optical dipole trap. The black curve is the power of the step potential beam. The red curve is the angle of the mirror directing the step potential beam. The green dotted line indicates the time of observation. (b) The sonic black hole, consisting of an elongated Bose-Einstein condensate of 8000 87Rb atoms. One of the 6000 repetitions of the spontaneous Hawking radiation experiment is shown. Due to the step potential, the flow of the atomic Bose-Einstein condensate is subsonic (supersonic) to the left (right) of the horizon. The horizontal and vertical black arrows indicate applied fluctuations. (c) The spectrum of the applied left/right fluctuations, corresponding to the horizontal arrows in (b). These fluctuations are added to the ramp of the red curve in (a). The dashed curve is the broadband fluctuations, whereas the solid curve is the filtered fluctuations. (d) Pairs of waves with positive correlations and constructive interference at the horizon. (e) Pairs of waves with negative correlations and destructive interference at the horizon. (f) The spectrum of correlations. The solid curve shows $S_0^2(|\beta|^2 + 1)|\beta|^2$, which corresponds to spontaneous Hawking radiation, whereas the dashed curve shows $S_0^2|\beta|^4$, which corresponds to classical fluctuations. The factor $S_0^2$ is included because it prevents the curves from diverging at low $k$.

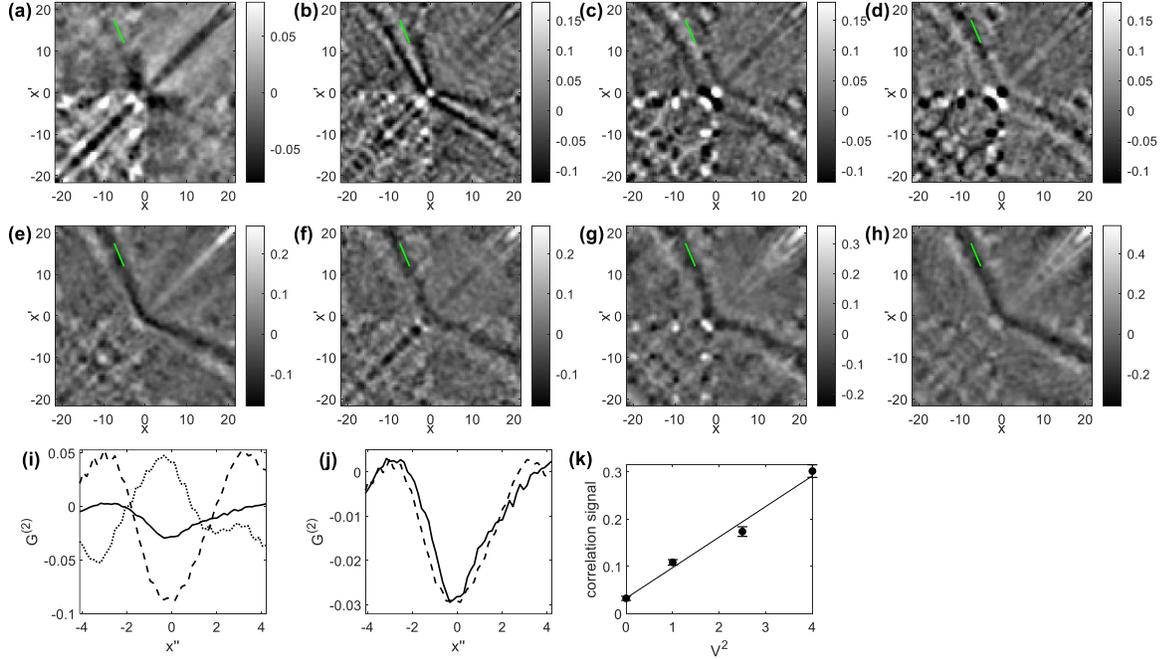

**Figure 2. A vibrating horizon generating Hawking-like radiation.** The density-density correlation function is shown. Positive (negative) values of $x$ or $x'$ are points inside (outside) the one-dimensional sonic black hole shown in Fig. 1(b). (a) Spontaneous Hawking radiation. (b) Broadband left-right vibrations. (c) Filtered left-right vibrations. (d) The negative of (c). (e) Slope vibrations. (f)-(h) Filtered slope vibrations with magnitudes of 0.5 V, 1.58 V, and 2 V RMS before filtering, respectively (The height of the step potential corresponds to 4 V). (i) The profiles of the correlation functions, from (a) (solid), (b) (dotted) and (e) (dashed). (j) the spontaneous (solid) and induced (dashed) profiles from (i). The dashed curve has been scaled vertically to allow comparison of the widths. (k) The magnitude of the signals from (a) and (f)-(h), which is the difference between the maximum and the minimum in the profiles. The $V^2 = 0$ value is the spontaneous signal. The line is a linear fit. The error bars indicate the standard error of the mean.

We will induce fluctuations of various types, and compare with Fig. 2(a). As a first attempt, we cause the horizon to move left and right, as indicated by the horizontal arrows in Fig. 1(b). This is achieved via small fluctuations in the ramp of the red curve in Fig. 1(a) (in this case, the fluctuations in the black curve are not used). The motion is random with an amplitude of 0.5 μm, and a broad temporal spectrum as indicated by the dashed curve of Fig. 1(c). This generates Hawking and partner-type particles, as seen by the correlation pattern of Fig. 2(b). The pattern however, contains short-wavelength fringes not seen for the actual Hawking radiation of Fig. 2(a). These narrow fringes reflect high frequencies in the vibrations of the horizon. Apparently, the frequencies emitted by the horizon are not limited by the Hawking temperature, when this technique is used. We thus filter the temporal spectrum with a width on the order of the Hawking temperature, as indicated by the solid curve of Fig. 1(c). The amplitude is 1.2 μm and 0.26 μm before and after filtering, respectively. The resulting correlation pattern shown in Fig. 2(c) is similar to the Hawking radiation of Fig. 2(a), with an important difference; the correlations are positive rather than negative. The same is true for Fig. 2(b) and its profile shown in Fig. 2(i) (dotted curve). Fig. 2(d) shows the negative of Fig. 2(c), and it indeed looks similar to Fig. 2(a). We can understand this qualitatively in the following way. Pairs of waves with positive correlations are illustrated in Fig. 1(d), whereas waves with negative correlations are illustrated in Fig. 1(e). At the horizon, there is constructive (destructive) interference for positive (negative) correlations, resulting in a large (small) density oscillation at the horizon. By causing the horizon to move left and right, we cause a large density oscillation at the horizon, which naturally generates positive correlations. Negative correlations are important because they correspond to positive mutual information between the inside and outside of the analogue black hole [29]. The Hawking radiation carries away entropy, leaving positive mutual information.

Judging from the relative strengths of the spontaneous and induced correlations in Figs. 2(a) and 2(d), respectively, the minuscule fluctuations required to produce the spontaneous quantity of Hawking radiation is on the order of 0.2 µm. This is even less than $\xi = 1.8$ µm, which is analogous to the Planck length, since it is the minimum length scale of the condensate wavefunction. Energies above this scale have a single particle nature, in analogy with quantum gravitational particles forming spacetime [30]. Indeed, 0.2 µm is less than the 0.6 µm interatomic spacing.

In order to achieve negative correlations, the location of the horizon should remain fixed despite the fluctuations. This can be achieved by causing fluctuations in the slope of the horizon, which corresponds to fluctuations in the Hawking temperature $T_\mathrm{H} = \frac{\hbar}{2\pi k_\mathrm{B}}(dv/dx - dc_\mathrm{s}/dx)|_{x=0}$. We cause this by inducing fluctuations in the height of the potential step as seen in the black curve of Fig. 1(a), and as indicated by the vertical arrows in Fig. 1(b). A very broad temporal spectrum (flat up to 45 $T_\mathrm{H}$) is applied. The resulting correlation pattern (Fig. 2(e)) exhibits the desired negative correlations, like the Hawking radiation pattern of Fig. 2(a). However, there is a difference between the spontaneous Hawking radiation and the induced Hawking-like radiation; the former has a narrower correlation band. This is also seen in the profiles of the bands (Fig. 2(j)), since the solid curve has a narrower minimum. We will now see that the narrower band reflects the quantum origin of the spontaneous Hawking radiation, as opposed to the fluctuations which we have added, which cannot possibly produce entanglement due to their deterministic nature [31].

The Fourier transform of each curve of Fig. 2(j) is proportional to $S_0\langle\hat{b}_\mathrm{H}\hat{b}_\mathrm{P}\rangle$, where $\hat{b}_\mathrm{H}$ and $\hat{b}_\mathrm{P}$ are annihilation operators for Hawking and partner particles, respectively [23, 32]. The $k$-dependent factor $S_0$ is common to both curves. Entanglement is determined by evaluating $\Delta = \langle\hat{b}_\mathrm{H}^\dagger\hat{b}_\mathrm{H}\rangle\langle\hat{b}_\mathrm{P}^\dagger\hat{b}_\mathrm{P}\rangle - |\langle\hat{b}_\mathrm{H}\hat{b}_\mathrm{P}\rangle|^2$, where $\Delta < 0$ implies entanglement [33]. In other words, if the correlations expressed by $|\langle\hat{b}_\mathrm{H}\hat{b}_\mathrm{P}\rangle|^2$ are larger than the product of populations $\langle\hat{b}_\mathrm{H}^\dagger\hat{b}_\mathrm{H}\rangle\langle\hat{b}_\mathrm{P}^\dagger\hat{b}_\mathrm{P}\rangle$, the state must be entangled. In the experiment where we impose classical fluctuations, the waves on either side of the horizon are caused by the same horizon motion, so the two waves should feature perfect classical correlations, which corresponds to $\Delta = 0$. Thus, $|\langle\hat{b}_\mathrm{H}\hat{b}_\mathrm{P}\rangle|^2 = \langle\hat{b}_\mathrm{H}^\dagger\hat{b}_\mathrm{H}\rangle\langle\hat{b}_\mathrm{P}^\dagger\hat{b}_\mathrm{P}\rangle$ in this case, and the Fourier transform of the induced correlation band gives a measure of $S_0^2\langle\hat{b}_\mathrm{H}^\dagger\hat{b}_\mathrm{H}\rangle\langle\hat{b}_\mathrm{P}^\dagger\hat{b}_\mathrm{P}\rangle$. Therefore, when we compare the widths of the spontaneous and induced correlation bands in Fig. 2(j), we are essentially comparing the inverse widths of $S_0^2|\langle\hat{b}_\mathrm{H}\hat{b}_\mathrm{P}\rangle|^2$ and $S_0^2\langle\hat{b}_\mathrm{H}^\dagger\hat{b}_\mathrm{H}\rangle\langle\hat{b}_\mathrm{P}^\dagger\hat{b}_\mathrm{P}\rangle$. The importance of these widths is evident in Fig. 1(f), which shows the theoretical curves for spontaneous Hawking radiation. Specifically, $S_0^2|\langle \hat{b}_H \hat{b}_P \rangle|^2 = S_0^2(|\beta|^2 + 1)|\beta|^2$, where $\beta$ is the Bogoliubov coefficient given by $|\beta|^2 = [\exp(\hbar\omega/k_B T_H) - 1]^{-1}$, while $S_0^2 \langle \hat{b}_H^\dagger \hat{b}_H \rangle \langle \hat{b}_P^\dagger \hat{b}_P \rangle$ is given by $S_0^2|\beta|^4$ [23]. The $S_0^2|\langle \hat{b}_H \hat{b}_P \rangle|^2$ solid curve in Fig. 1(f) is wider (broader in frequency), and therefore lies above the $S_0^2 \langle \hat{b}_H^\dagger \hat{b}_H \rangle \langle \hat{b}_P^\dagger \hat{b}_P \rangle$ dashed curve, which implies entanglement. Thus, the narrower width of the spontaneous correlation band in Fig. 2(j) reflects entanglement.

In contrast to the broad temporal spectrum of Fig. 2(e), Figs. 2(f)-(h) show an applied spectrum with a width on the order of the Hawking temperature. Since the two cases are similar, it is seen that the filtering has essentially no effect. It seems that the slope of the horizon is a natural spatial filter which gives the Hawking spectrum, so the temporal filtering is unnecessary for the case of height fluctuations. Fig. 2(k) shows the magnitude of the signals from Fig. 2(a) and Figs. 2(f)-(h), as a function of the magnitude squared of the vibrations (the variance of the black curve in Fig. 1(a)). It is seen that the Hawking radiation generated is proportional to the magnitude squared of the horizon vibrations. We will see that the same relationship should apply to real black holes.

**Quantum fluctuations of real black hole horizons in terms of quasinormal modes**

In this section we will see that quantum fluctuations of the horizon of a real black hole, described in terms of the fundamental quasinormal modes, provide a much better description of Hawking radiation than does the usual blackbody model. The fundamental quasinormal mode for gravitons is illustrated in Fig. 3(a). Its frequency has real and imaginary parts $\omega_R$ and $\omega_I$ given by [7, 8]

$$\omega_R + i\omega_I \equiv 4\pi(\Omega_R + i\,\Omega_I)\frac{k_B T_H}{\hbar} \qquad (1)$$

where $\Omega_R = 0.747$ and $\Omega_I = 0.178$.

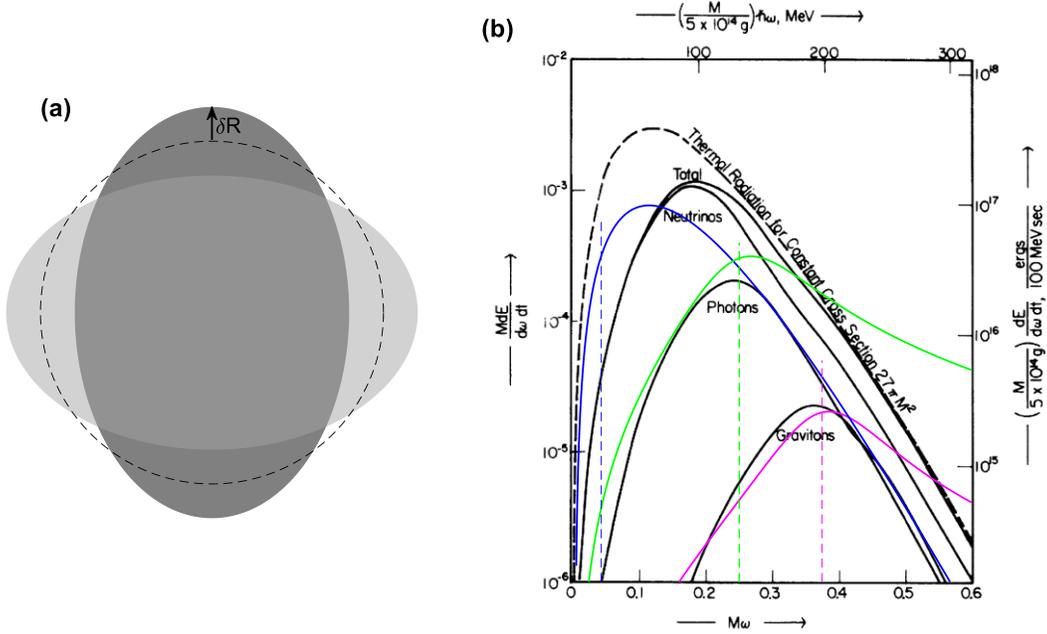

**Figure 3. The Hawking spectrum compared with the fundamental quasinormal mode.** (a) Illustration of the fundamental quasinormal mode. The horizon oscillates between the dark gray and light gray ellipses. The dashed curve is the unperturbed spherical horizon. (b) The power spectrum of Hawking radiation including graybody factors, taken from Ref. 10. The relevant curves are labeled "Gravitons" and "Photons". We have added the colored curves. The blue curve indicates the Hawking blackbody spectrum of gravitons or photons, with no graybody factor. The dashed blue line indicates the Hawking temperature. The solid magenta curve is the spectrum of dynamical Hawking radiation of gravitons, emitted by the fundamental quasinormal mode, with no adjustable parameters. The dashed magenta line indicates the frequency of the fundamental quasinormal mode. The green curve and dashed line correspond to photons, with no adjustable parameters.

As seen in Fig. 3(b) taken from Ref. 10, the peak in the Hawking spectrum for gravitons (labeled "Gravitons") is shifted relative to the thermal blackbody form (blue curve), and encloses a much different area. This discrepancy is described in terms of graybody factors [10, 34]. The location of the graviton peak is much closer to the frequency of the fundamental quasinormal mode (dashed magenta line), than to the peak in the thermal blue curve. Furthermore, the spectrum of the fundamental quasinormal mode is broadened due to damping, as indicated by the solid magenta curve. Since the solid magenta curve matches the graviton curve fairly well with no adjustable parameters, we see that only one quasinormal mode is sufficient to emit the entire Hawking spectrum of gravitons.

Fig. 3(b) also shows a discrepancy for Hawking radiation of photons, since the "Photons" curve is shifted relative to the blue thermal curve. There is also a fundamental quasinormal mode which emits photons rather than gravitons, with a frequency given by $\Omega_R = 0.497$ and $\Omega_I = 0.190$ [35]. The real part of the frequency is indicated by a green dashed line in Fig. 3(b), which agrees well with the center of the "Photon" curve. The solid green curve indicates the spectrum of the fundamental quasinormal mode, which agrees fairly well with the "Photon" curve, with no adjustable parameters.

There are two ways to obtain the solid magenta and green curves, each reflecting quasinormal mode fluctuations of different natures. Firstly, one can simply compute the Fourier transform $A(\omega)$ of a damped sinusoid at the quasinormal mode frequency. The power radiated $dP/d\omega$ is proportional to $\omega^2|A(\omega)|^2$, which gives

$$\frac{dP}{d\omega} \propto \frac{\omega^2}{(\omega^2-\omega_R^2+\omega_I^2)^2+4\omega_I^2\omega_R^2} \qquad (2)$$

Furthermore, we demand that the total power in the spectrum be

$$P = 2\omega_I \frac{\hbar\omega_R}{e^{\hbar\omega_R/k_B T_H}-1} \qquad (3)$$

where the second factor (the ratio) is the energy of a mode at the Hawking temperature $T_H$, and $2\omega_I$ is the rate of energy decay in the quasinormal mode. Eq. 2 normalized by Eq. 3 gives the magenta and green curves of Fig. 3(b). In this first picture, the quantum fluctuations of the horizon are tiny ringdowns which are occasionally excited, and subsequently decay, emitting a burst of Hawking radiation each time. In the second picture, we demand a continuous emission. In order to obtain it, we imagine that the quasinormal mode is driven by an effective flat spectrum, which might be associated with quantum fluctuations. Modeling this by a damped harmonic oscillator driven with frequency $\omega$, the power dissipated is given by Eq. 2, so the second picture also gives the magenta and green curves in Fig. 3(b).

We have found that a single quasinormal mode is sufficient to emit the entire spectrum of Hawking radiation. Nevertheless, we should consider the contribution of the other quasinormal modes. The higher harmonics have higher real frequency [8], so they would not contribute significantly by Eq. 3. The overtones of the fundamental mode have lower real frequency, but they also have much stronger damping, so they would have a much weaker emission for the same effective driving.

**The amplitude of the horizon fluctuations**

The radiated power of a quasinormal mode is proportional to the amplitude squared of the horizon fluctuation. It seems reasonable that the same proportionality factor should apply to small but classical fluctuations, as well as very small quantum fluctuations. This is reminiscent of dipole emission by accelerating electrons, where the radiated power is proportional to the amplitude squared of the motion, with the same proportionality factor for radio antennas and Rayleigh scattering by atoms. It is also reminiscent of a harmonic oscillator, for which the energy is proportional to the amplitude squared, with the same proportionality factor for both large amplitude classical oscillations and quantum, zero-point motion. Assuming that the proportionality factor is indeed the same for both ringdowns and Hawking radiation of gravitons, one obtains

$$\frac{P_H}{\delta R_H^2} = \frac{P_{RD}}{\delta R_{RD}^2} \qquad (4)$$

where the amplitude $\delta R$ of the horizon fluctuations is illustrated in Fig. 3(a). The left side of Eq. 4 corresponds to Hawking radiation of gravitons, and the right side corresponds to the ringdown. In this section, we will use Eq. 4 to find the amplitude $\delta R_H$ of the quantum fluctuations of the horizon. To that end, we will need to evaluate the other quantities in the equation.

Firstly, we will evaluate $P_H$. When viewed from infinity, a black hole has a large apparent radiating area of $27\pi R^2$ [36]. When traced back to the black hole, all rays emanate from the horizon, so the large apparent area does not imply that Hawking radiation emanates from a quantum atmosphere [37, 38]. Due to the large area, the total power in Hawking radiation should be $27\pi R^2 \sigma T_H^4$, where $\sigma$ is the Stefan-Boltzmann constant [10]. This power corresponds to the area under the blue curve of Fig. 3(b). On the other hand, the area under the magenta curve, which is 0.04 of the area under the blue curve, provides a more accurate estimate of the area under the "Graviton" curve. Thus, we will use Eq. 3 as $P_H$. Furthermore, we will take the Hawking temperature to be $T_H = \hbar c / 4\pi k_B R$, where $R$ is the radius of the horizon, and we have ignored the spin of the black hole for simplicity, as we do throughout this work.

Now we will evaluate the right side of Eq. 4. This expression is classical, and can be evaluated by general relativistic considerations. In order to obtain fairly precise estimates which are relevant for actual black holes, we use the waveforms computed for black hole mergers. Figure 4(a) shows a numerical simulation of the strain caused by a black hole merger from Ref. 1, in which the parameters of the simulation were adjusted to match the experimental observation. We use the simulation rather than the measurement since the latter is too noisy for our purposes. We extract the real part of the frequency from the zero crossings, as shown in Fig. 4(b). The radiated power is proportional to the amplitude of the oscillations in Fig. 4(a) squared, times the frequency squared of Fig. 4(b), as shown in Fig. 4(c). The curve has been normalized so its integral is the total energy emitted. The peak luminosity occurs at the point $P_{\text{peak}}$, which is almost simultaneous with the strain peak for this merger. This point also marks the instant that the two black holes are separated by $1.4\,R$ [1], and have formed a single black hole which begins to ring down [1]. A separation of $1.4\,R$ corresponds to an amplitude $\delta R \approx 0.7\,R$. Thus, when evaluated at the peak, the ratio on the right side of Eq. 4 is approximately $P_{\text{peak}}/(0.7\,R)^2$. On the other hand, the fundamental quasinormal mode begins to dominate 6 ms later, when the frequency becomes a constant and the exponential decay in the power begins, as indicated by the highlighted sections of Figs. 4(b) and 4(d). At this later time, the power has decreased to $0.3\,P_{\text{peak}}$, and $\delta R$ has decreased also. Since we do not know $\delta R$ at this time, we merely use the ratio at the peak as an order-of-magnitude estimate. Thus, we obtain

$$\frac{P_H}{\delta R_H^2} \approx \frac{P_{\text{peak}}}{(0.7\,R)^2} \tag{5}$$

Equation 5 was obtained for a particular merger, but we use this approximate expression for all mergers.

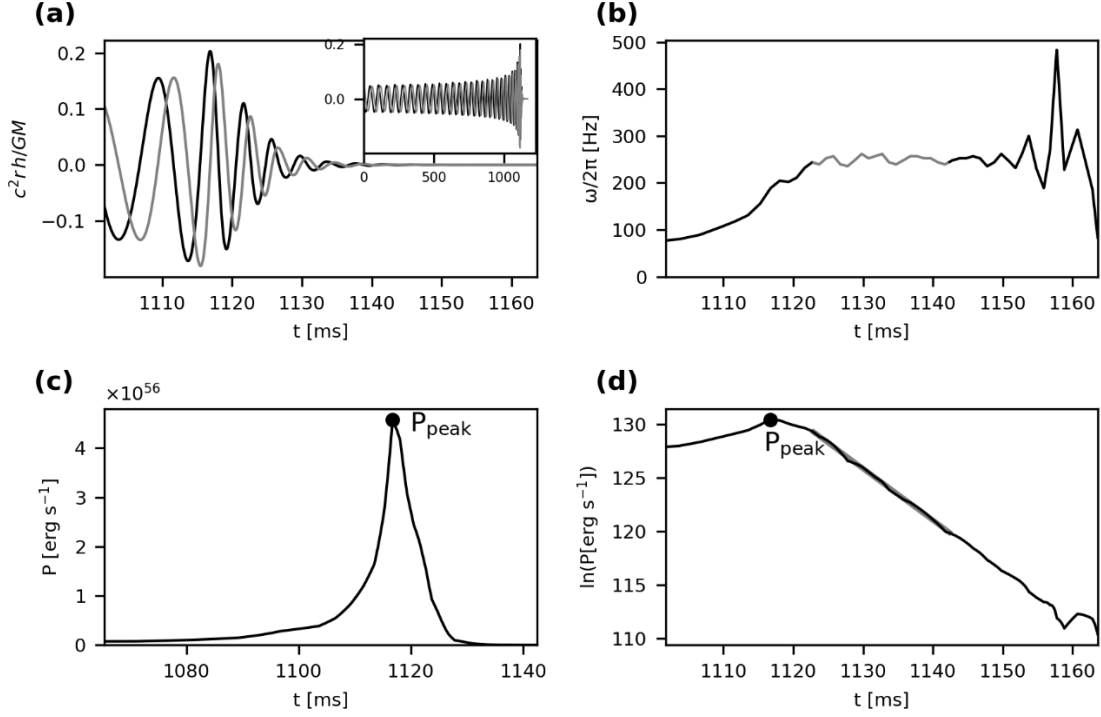

**Figure 4. Analyzing gravitational wave data.** A numerical simulation of black hole merger GW150914 is shown, from Ref. 1. (a) The strain. The black (gray) curve is the + (×) polarization. (b) The real part of the frequency. The gray section of the curve indicates the constant region where the fundamental quasinormal mode dominates. (c) The radiated power. The point $P_{\text{peak}}$ marks the peak luminosity. (d) The log of the power. The line is a linear fit.

Figure 5(a) shows the peak luminosity for all the black hole mergers for which there are published simulations (to our knowledge), including the events GW150914, GW151226, GW170104, GW190412, and GW190521 [1, 3-5, 39]. Each point is found like point $P_{\text{peak}}$ of Fig. 4(c). Figure 5(b) shows $\delta R_H$ predicted by each point in Fig. 5(a), by Eq. 5, in units of the Planck length $\ell_P = \sqrt{\hbar G/c^3}$. It is seen that the horizon fluctuations are on the order of 0.1 Planck lengths. Equation 5 can also be written as

$$\frac{\delta R_H}{\ell_P} \approx 0.7 \left(\frac{R}{\ell_P}\right) \sqrt{\frac{P_H}{P_{\text{peak}}}} \tag{6}$$

Thus, the $\delta R_H \approx \ell_P$ result is based on the similarity between two large numbers: The radius of the black hole in Planck units $R/\ell_P$, and the square root of the peak luminosity in Hawking-radiation units $(P_{\text{peak}}/P_H)^{1/2}$. These two numbers are shown in Fig. 5(d).

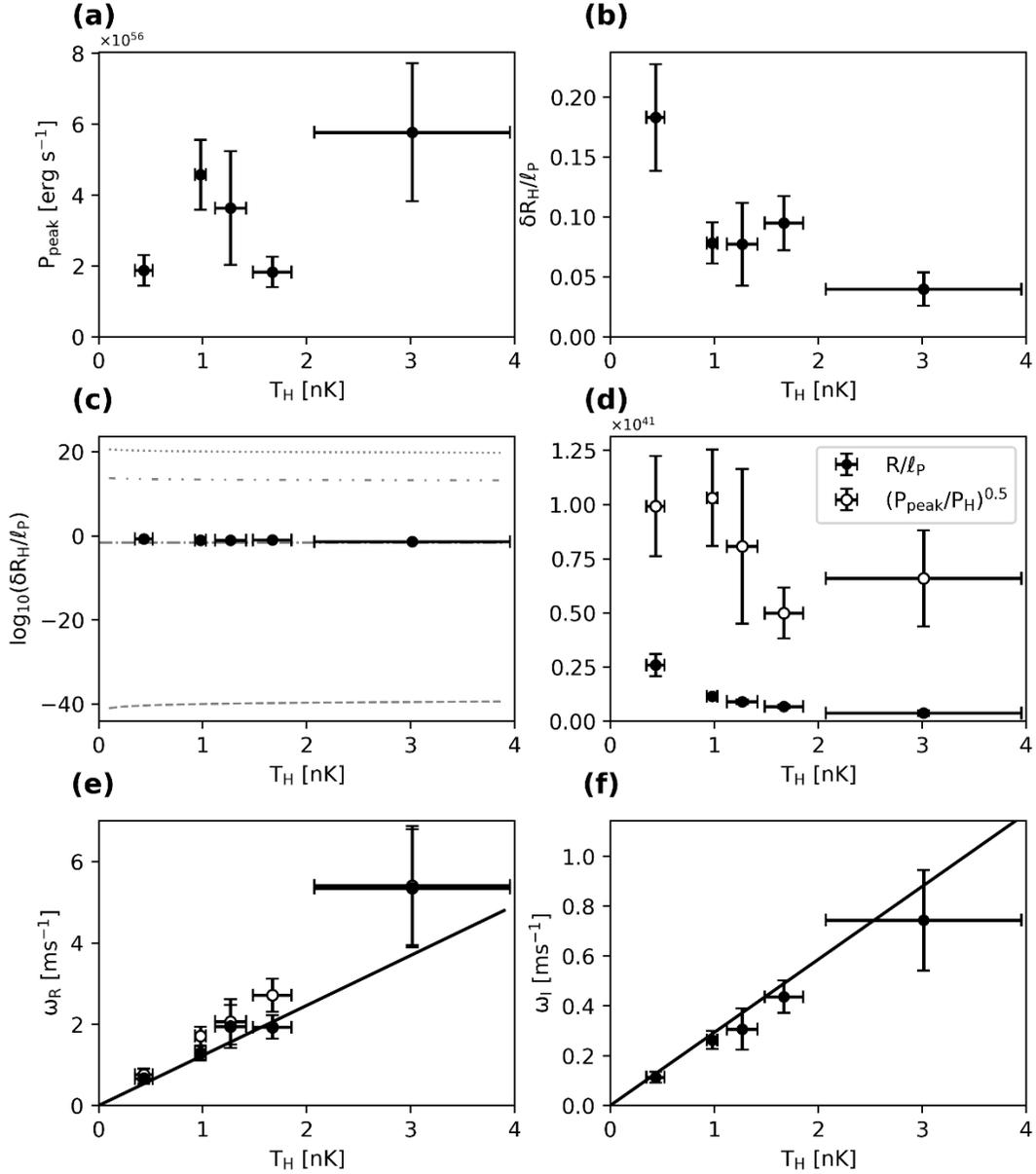

**Figure 5. The predicted quantum fluctuations of the horizon.** The error bars indicate 90% credible intervals. (a) Peak luminosity in black hole mergers. (b) The predicted $\delta R_H$. (c) The log of $\delta R_H$. The predictions from Refs. 11, 16, 17 and 18 are indicated by dashed, dash-dotted, dash-double dotted and dotted lines, respectively. (d) Filled circles indicate $R/\ell_P$. Open circles indicate $\sqrt{P_{\text{peak}}/P_H}$. (e) The real part of the frequency. The filled circles indicate the frequency at the time of the peak luminosity. The open circles are the average over the period where the fundamental quasinormal mode dominates, which is the gray section of the curve in Fig. 4(b). The line indicates Eq. 1. (f) The imaginary part of the frequency. Each point is half the slope of the linear fit in Fig. 4(d). The line indicates Eq. 1.

The various theoretical predictions are compared with $\delta R_H$ in Fig. 5(c). York's prediction (dash-dotted line) is very good. On the other hand, the semiclassical predictions based on Newtonian (dash-double dotted line) and relativistic (dotted line) calculations are far too high, while the prediction of Ref. 11 (dashed line) is far too low.

As a check that the ringdown is dominated by the least-damped fundamental quasinormal mode illustrated in Fig. 3(a), Figs. 5(e) and 5(f) compare the extracted values of the real and imaginary parts of the frequency $\omega_R$ and $\omega_I$ with the theoretical expression given by Eq. 1. Reasonable agreement is seen between the measured points and the black lines indicating Eq. 1, despite the overtones with higher damping, which should play a role at early times after the peak [40].

**Conclusions**

We have explored the possibility that Hawking radiation of gravitons and photons is emitted dynamically by quantum fluctuations of the horizon, in the form of quasinormal modes. This view emerges naturally, since the Hawking temperature is well below the lowest resonant modes. One thus obtains discrete, broadened spectral lines, corresponding to the fundamental quasinormal modes for gravitons and photons. These emission spectra are much more accurate than the usual Hawking blackbody spectrum. In other words, the Fourier transform of the ringdown gives the Hawking spectrum of gravitons. Our analogue experiment is consistent with the dynamical Hawking radiation picture, in that horizon vibrations are seen to produce Hawking/partner-like pairs. The experiment does not explore the entanglement since we apply deterministic vibrations which could not possibly produce entangled pairs [31]. Indeed, the spontaneous Hawking radiation is seen to be entangled, but the applied vibrations give a non-entangled result. Furthermore, the analogue experiment shows us the importance of the sign of the correlations between the inside and outside of the black hole, for discerning if radiation is truly of the Hawking type. It would be worthwhile to compare the signs of the predicted correlations for Hawking radiation and quasinormal modes in a real black hole. Also, we extrapolate from classical quasinormal mode oscillations to the quantum fluctuations of the horizon which emit Hawking radiation. The classical oscillations are readily evaluated, and we use simulations of ringdowns after black hole mergers for accurate estimates. We find that the quantum fluctuations of the horizon should be on the order of 0.1 Planck lengths, for the black hole mergers for which we have sufficient information. Furthermore, there may be other non-radiative degrees of freedom on the horizon which contribute to the entropy, in addition to the fluctuations studied here [41, 42]. This work brings Hawking radiation one step closer to quantum gravity, since horizon vibrations directly cause the emission.

We thank A. Ori, T. Jacobson, T. Hinderer, S. Liberati, R. Balbinot, B. L. Hu and U. Fischer for helpful comments. This work was supported by the Israel Science Foundation, grant 531/22.